\documentclass[%
reprint,
nofootinbib,
amsmath,amssymb,
aps,
floatfix,
]{revtex4-2}

\usepackage[normalem]{ulem}
\usepackage{graphicx}
\usepackage{dcolumn}
\usepackage{bm}

\usepackage{comment}
\usepackage{xcolor}

\newcommand{\etal}{\textit{et al.~}}

\newcommand{\vecr}{{\vec r}}

\begin{document}
	
	\preprint{APS/123-QED}
	
	\title{Pairing enhancement through the photography of the intermediate nucleus spectrum in a two-nucleon transfer process}
	
	\author{G.~Singh}
	\email{gsingh@uni-mainz.de}
	\affiliation{Institut f{$\ddot{u}$}r Kernphysik, Johannes Gutenberg-Universit{$\ddot{a}$}t Mainz, 55099, Mainz, Germany}
	
	

	\date{\today}
	

\begin{abstract}
		
		
		
		
		\textbf{Abstract:} While forming an ($A$+2) nucleus from a nucleus $A$ via a two-neutron transfer reaction, the constructive interference of the many possible reaction channels favors significant pairing enhancement through the continuum of the intermediate ($A$+1) nucleus [Phys. Lett. B \textbf{834} 137413 (2022)]. I analyse this situation in more generality, from the point of view of a varying pairing field and different continua leading to the formation of ($A$+2) nucleus. I consider $^6$He and $^{22}$C, described as housing two-neutrons in orbitals of $^5$He and $^{21}$C, respectively. 
		The different possible situations manifest that the continuum correlations are extremely crucial to the extension of the pairing enhancement observed in the system.
\end{abstract}
	
	\maketitle
	

\section{Introduction}
\label{sec:intro}

Transfer reactions in nuclei are generally sensitive to the tail of the wave functions. This property is widely used to study them using peripheral reaction techniques like the asymptotic normalization coefficient (ANC) method \cite{MSB17PRC}. The overestimation of this tail, however, due to interaction descriptions via standard shell model potentials push for a systematic study of absolute cross-sections due to such transfers \cite{BWBook}. These studies are almost well understood for celebrated stable nuclei with high binding energies per nucleon (viz.,$^{12}$C, $^{16}$O, etc.), but become non-trivial when considering loosely bound exotic nuclei and open quantum systems. A significant fraction of the non-triviality and the ensuing challenges in the elucidation of reaction mechanisms is due to nuclear correlations - a consistently encountered lineament throughout the nuclear chart. Pairs of identical particles are linked by correlations that play vital roles in describing nuclear structure, influencing phenomena from fission to neutron star equation of state (via symmetry energy) to magic numbers (shell effects) to the existence of halos \cite{SNS16PRC,Tanaka21Science,HS05PRC,PPP15SSR,Brown13PRL}. These correlations have striking regularities which shape nuclear spectra and can result in transition strengths that are much higher than single particle estimates.

In nuclei near the drip lines, the weakly bound nature of the valence particles forces the existence of diffused halos and scatters the so formed Cooper pairs (that behave as an isolated system in coordinate space) into the continuum. This can highly affect the character of the short ranged pair correlations \cite{DNW96PRC,OV01RPP,PIB13RPP}. In case of weakly bound nuclei on the proton rich side, the continuum coupling effects are much less crucial due to the Coulomb barrier. For weakly bound neutron systems, the pair fields furnish consequential couplings between neutron pairs in the bound states and neutrons kinetic in the low-energy continuum \cite{Hama03PRC}. These neutron correlations are known to stabilize Borromean nuclei and have been the subject of a number of studies \cite{MMS05PRC,HSS10JPG,Yama22PRC}.


Such correlations and pairing dynamics, manifested also by vibrational and rotational couplings, can be easily analysed by two-nucleon transfer reactions. Via dineutron captures on $\alpha$ particles and He isotopes, two-neutron transfers have also been instrumental in navigating through the $A$ = 5 and $A$ = 8 mass gaps. In fact, these transfers are so important that they are known to have some effect on the low-temperature $r$-process seed nuclei production in a neutron rich environment \cite{BGM06PRC,GHT95PRC}. While one-neutron transfers, so far, have been quite interesting in explaining the features of weakly bound one-neutron halo nuclei in the lower mass region \cite{MN06NPA,SD19PRC,PLM23PRC,CGM17PLB,GMG15PRC,YC18PRC,MYC19PRC}, two-neutron transfers help us understand the stability of Borromean nuclei away from the valley of stability \cite{TAB08PRL,PIB13RPP,PBV10PRL,Keeley07PLB,SFV23JPCSr,BBP19EPJ,SFV22PLB,LFV14PRC,PIB14NPN,TSK13PPNP,ECL16PRC,CKC19PLB}. 

It is well-established that pairing correlations and the couplings between the states of the intermediate ($A$+1) nuclei feature to enhance the cluster and/or two-particle transfers \cite{PIB13RPP,OV01RPP,VMH15AIP,HS16FBS,HS05PRC}. This `pairing enhancement' originates from the coherent interferences of different paths through the states in the ($A$+1) nuclei that arise due to the correlations in the initial and final state wave functions	\cite{OV01RPP,LFV14PRC}. These correlated 
wave functions hold the requisite microscopic and nuclear structure information to extract two particle transfer amplitudes. This gives the weight of each of the two-step paths, which in turn can be obtained from single particle transfer form factors.
	 
For weakly bound as well as exotic nuclei, the various paths which interfere coherently and contribute to this pairing enhancement, almost all pass through the continuum which makes it imperative to include the continuum in transfer calculations \cite{VP10NPA,PIB13RPP,OV01RPP,MN06NPA,VMH15AIP,MMV21PRC,Betan17NPA,UDN12PRC,HOM22PPNP}. Although this boosts the transfer to the final state of the final product \cite{SFV22PLB}, however, the inclusion of the intermediate continuum is not so straightforward and can be quite demanding, as one might lose out on convergence and orthonormality conditions due to the non square-integrable nature of the continuum wave functions.

An assessment of such a two-neutron transfer proccess, including the continuum coupling effects was recently done in Ref. \cite{SFV22PLB}. The study of sequential transfers indicated a decreasing probability contribution for the pairs in higher states of the continuum. This was, of course, offset when there was the presence of a resonance. When computing two-neutron transfer cross-sections to the experimentally known ground state (g.s.) of $^6$He, through an unbound $^5$He compared to a hypothetically bound intermediate $^5$He, it was seen that there was significant pairing enhancement through the former case. This presented a rather cute way for an ($A$+2) nucleus to be bound, especially in stellar plasma or astrophysical scenarios, despite the ($A$+1) being unbound.

In the present work, I try and see this enhancement in more generality, specifically when the final state of $^6$He is not fixed, courtesy of a varying pairing field. Three cases will be considered for three different perspectives, where the intermediate $^5$He will be bound, just bound and unbound. The first two scenarios are depicted by imagining $^5$He with one neutron separation energy, $S_n$ = 1\,MeV and 0.1\,MeV, respectively. These act as our control. The naturally occuring unbound $^5$He is taken with its resonance generated at 0.69\,MeV above zero. Ideally, one should expect pairing enhancement to any of the states in $^6$He, but the degree of enhancement through the photography of the spectrum of the intermediate nucleus needs to be quantified for each of these hypothetical final states in $^6$He. Therefore, I study a two-neutron transfer via the reaction $^{18}$O($^4$He,$^6$He)$^{16}$O. In addition, the $^{18}$O($^{20}$C,$^{22}$C)$^{16}$O reaction is also analysed, with a view to study a similar parity case as well provide a scenario where there is only a virtual state present in the intermediate nucleus ($^{21}$C). This part is composed of two cases where I vary the contribution an unperturbed continuum provides, so as to see the effect on the enhancement due to the non-resonant continuum portion. 

Further, in the present analysis, unlike the one in Ref. \cite{SFV22PLB}, I stick to the high beam energy scenario as it is always better to populate more energy states in the continuum. I then compute the two-neutron transfer cross-section sequentially for all the cases under scrutiny and eventually see the effect of pairing enhancement via the different final states in the final product $^6$He as well as different unperturbed continuum contributions in $^{21}$C.

In the next section, I briefly discuss the wave functions required for this study and after dicussing the results in Section \ref{sec:results}, we see the conclusions in Section \ref{conclusion}.
	
	
\section{Description of the $^5$H\lowercase{e} \& $^{21}$C spectra}
\label{sec: formalism}
	
{I start by considering the two-neutron transfer reactions $^{18}$O($^4$He,$^6$He)$^{16}$O and $^{18}$O($^{20}$C,$^{22}$C)$^{16}$O, where I assume that the lab beam energy of the $^{18}$O projectile is 100\,MeV. Such a high beam energy enables one to populate the higher lying states in the continuum, addtionally providing the relief of not having to worry about the possible effects of the $Q$-value \cite{SFV22PLB,OV01RPP}. Further, since the projectile and target nuclei, all have spin-parity of $0^+$, choosing a sequential transfer study ensures that there is no contribution of the simultaneous transfer, as it gets cancelled by the non-orthogonality terms in the \textit{prior-prior} or \textit{post-post} picture \cite{BWBook, PIB13RPP,SFV22PLB}. Therefore, I work in the \textit{prior-prior} representation for this sequential two-neutron transfer. For the purpose of calculations, a modified version of the Transfer Form Factors (TFF) code \cite{TFF} was used.}
	
The first step then for the first reaction, obviously, is the transfer of a single neutron resulting in the formation of $^5$He and leaving an intermediary complement of $^{17}$O. For simplicity, any configuration mixing is neglected in this analysis and $^{17}$O is supposed to be composed of just the 1$d_{5/2}$ component to its ground state, bound by 4.1433\,MeV \cite{Wang17CPC}. In $^5$He, only the $p_{3/2}$ level in the bound as well as the continuum states receives the transferred neutron. The aim of these approximations is to prevent complications due to the mixing of different angular momenta. Any excitations or de-excitations in either $^5$He or $^{17}$O are also avoided. However, for the $^{18}$O($^{20}$C,$^{22}$C)$^{16}$O reaction, the intermediary $^{17}$O, in its ground state, was composed of both the 1$d_{5/2}$ component (64\%) as well as the 2$s_{1/2}$ component (36\% at -3.2726\,MeV \cite{Wang21CPC}), while the $^{21}$C continuum is supposed to house only the $s_{1/2}$ levels.


{The theory of successive neutron transfers, as is very well known, is deeply dependent on the form factors involving the intermediate ($A+1$) nuclei (cf. Eq. (44) on page 423 of Ref. \cite{BWBook} and also Ref. \cite{OV01RPP}). These form factors need to be computed for each of the states that describe the picture of the energy spectrum of $^5$He and $^{21}$C. Since they both are naturally unbound, ideally one should be able to describe their entire energy spectrum as a continuum. However, as averred, the inclusion of the continuum in reaction calculations is a very arduous task. Therefore, for realistic purposes, one needs to discretise this continuum into a finite number of states resulting in meaningful calculations.}

To obtain the eigen functions corresponding to these discretised states, I employ the so called pseudo-state (PS) method, consisting of diagonalizing the Hamiltonian in a finite basis of square integrable functions. For the present purpose, I chose the transformed harmonic oscillator (THO) basis, which helps describe the wave functions of a system by expanding them via,

	\begin{equation}
	\psi^{jm_j}(\vec r,\Omega) = \sum_{\beta}\sum_{i}^{i_{max}}C_{i\beta}^{j}{\cal R}_{\beta}^{j,THO}(r)\mathcal{Y}_{\beta}^{jm_j}(\hat r,\Omega),
	\label{eq:3bwf}
\end{equation}

where $\mathcal{Y}_{\beta}^{jm_j}(\hat r,\Omega)$ provide the angular wave functions and are given by,

\begin{equation}
\mathcal{Y}_{\beta}^{jm_j}(\hat r,\Omega)= \left[\varUpsilon_{ls}^{J}(\hat r)\otimes\chi_{I}(\Omega)\right]_{jm_j}.
	\label{eq:Upsilon0}
\end{equation}

Here, $j$ is the total angular momentum (with projection $m_j$) that is a result of the coupling between core angular momentum $I$ and angular momentum $J$ of the valence neutron. Of course, $J$ is the resultant of spin of valence neutron $s$ coupling to the orbital angular momentum $l$ relative to the core. Then, $\beta$ denotes a set of quantum numbers $\{l,s,J,I\}$ involved corresponding to each channel. The resultant wave function due to the coupling of the valence neutron with the corresponding spherical harmonic is described by $\varUpsilon_{ls}^{J}(\hat r)$, while $\Omega$ denotes the core degrees of freedom.

The rest of the expression in Eq. (\ref{eq:3bwf}), $i.e.$, $\sum_{i}^{i_{max}}C_{i\beta}^{j}{\cal R}_{\beta}^{j,THO}(r)$ gives the square integrable radial wave functions under the THO scheme. $C_{i\beta}^{j}$ are the expansion coefficients that can easily be obtained by diagonalizing the Hamiltonian for basis size $i$. The wave functions are called transformed because they are derived as a result of a transformation from harmonic oscillator (HO) functions via,

\begin{equation}
	\label{eq:tho}
	{\cal R}^{THO}_{n, \l}(r)= \sqrt{\frac{du}{dr}} R^{HO} _{n, \l}[u(r)].
\end{equation}
	%

	


Here, $R ^{HO}_{n, \l}(s)$ (with the principal quantum number $n=1,2,\ldots$) is the radial part of the HO functions while $u(r)$ defines the transformation that is used to go from the HO basis to the THO basis. This transformation is called the local scale transformation (LST) and is given analytically by Karataglidis \etal \cite{KAG05PRC} as,
	\begin{equation}
	u(r) = \frac{1}{b\sqrt{2}}\left[\frac{1}{\left(\frac{1}{r}\right)^{4} +
		\left(\frac{1}{\gamma\sqrt{r}}\right)^4}\right]^{\frac{1}{4}}.
	\label{eq:LST}
\end{equation}
	
Here, $\gamma$ is related to the transition radius and $b$ is called the oscillator strength. It is crucial that to generate the wave functions for discretised states, optimum values of these parameters are used and this is where LST is a rather useful approach. It not only converts the Gaussian asymptotic behaviour of the HO functions to a simpler exponential form ($e^{-\gamma^2r/2b^2}$), thereby facilitating convergence (which is highly desirable when dealing with continuumm states), it also allows one to control the density of pseudo-states near the threshold via the $\gamma/b$ ratio. A smaller ratio of $\gamma/b$ results in a larger density of states near the continuum threshold, a desirable property for weakly bound nuclei since they have only one or two bound states. In fact, for a given $b$, as $\gamma$ decreases, the basis functions tend to be more asymptotic in their behaviour. The negative eigenvalue wave functions are then identified with the energies of the bound states, whereas the positive ones provide a discrete representation of the continuum. More details about these parameters as well as about THO basis in general can be found in Refs. \cite{LMA10PRC,LMA12PRC,LMA14PRC,SSC22PRC,CRA13PRC,Casal18PRC,CSF20PRC}. The THO wave functions are just one of many available wave functions and one is free to use other approaches as well \cite{MPV16JPG}.

In the present analysis, the values of $\gamma$ and $b$ were fixed at 1.8, 2.0, 2.0 and 1.0, 1.0, 1.2\,fm for the three cases of $^5$He nucleus, i.e., bound with $S_n$= 1\,MeV, bound with $S_n$= 0.1\,MeV and unbound, respectively \cite{SFV22PLB}. For $^{21}$C, the values were fixed at 2.0 and 1.2 for $\gamma$ and $b$, respectively. These values enabled an optimum spatial and density representation of the discretized continua for a given energy range \cite{LMA10PRC,MAG09PRC}.


For the discretised energy states in $^5$He, the binding energies of the respective bound cases were reproduced at 1\,MeV and 0.1\,MeV below zero by adjusting the depth of the potential well in a single particle picture. They were the only bound states in each case while a set of discretised states imitated the continuum. For the unbound case, the THO discretised continuum states were so adjusted that the third state generated in the spectrum, at 0.69\,MeV above zero, was closest to the resonance in $^5$He (which is at 0.735\,MeV for the $p_{3/2}$ state \cite{Wang21CPC}). This was done to best reproduce the natural conditions available in $^5$He. In all, there were a total of 8 states (that yielded convergent results) that were used to successfully describe the $^5$He spectra in each of the cases \cite{SFV22PLB}. A similar set of 8 states were also used to discretise the $^{21}$C continuum. There is a virtual state in the system \cite{Jagjit19FBS} and this was mimicked as the first discretised state at an energy of 0.0288\,MeV.

For the purpose of clarity, the next section discusses the Helium results first and then a transition to Carbon is made separately.

\section{Results and Discussion}
	\label{sec:results}
	
\subsection*{One-neutron transfer}
	
For a sequential two-neutron transfer, one must analyse first the single neutron transfer. This requires the calculation of single partcle form factors. 
Once the spectra are generated for the various cases of interest and the form factors are calculated in the \textit{prior} picture (using Eq. (19b) on page 330 of Ref. \cite{BWBook}), the transfer cross-sections can easily be calculated. Care must, however, be taken in choosing the \textit{prior} or the \textit{post} form, for a wrong choice would lead to form factors with asymptotic oscillations and unphysical extensions. These extensions are beyond the normal range of elongations that one would expect with form factors\footnote{The form factors were found to be quite non-local with a long range of overlap of the wave functions, but this was due to the weakly bound nature of the nuclei rather than the choice of the \textit{prior} or \textit{post} picture. {This just makes the convergence harder to be achieved.}} for weakly bound nuclei \cite{OV01RPP} and would result in enormous transfers \cite{SFV22PLB}. 

\begin{figure}[htbp] 
	\centering
	\includegraphics[trim={0 0 0 0},clip,width=1.0\columnwidth]{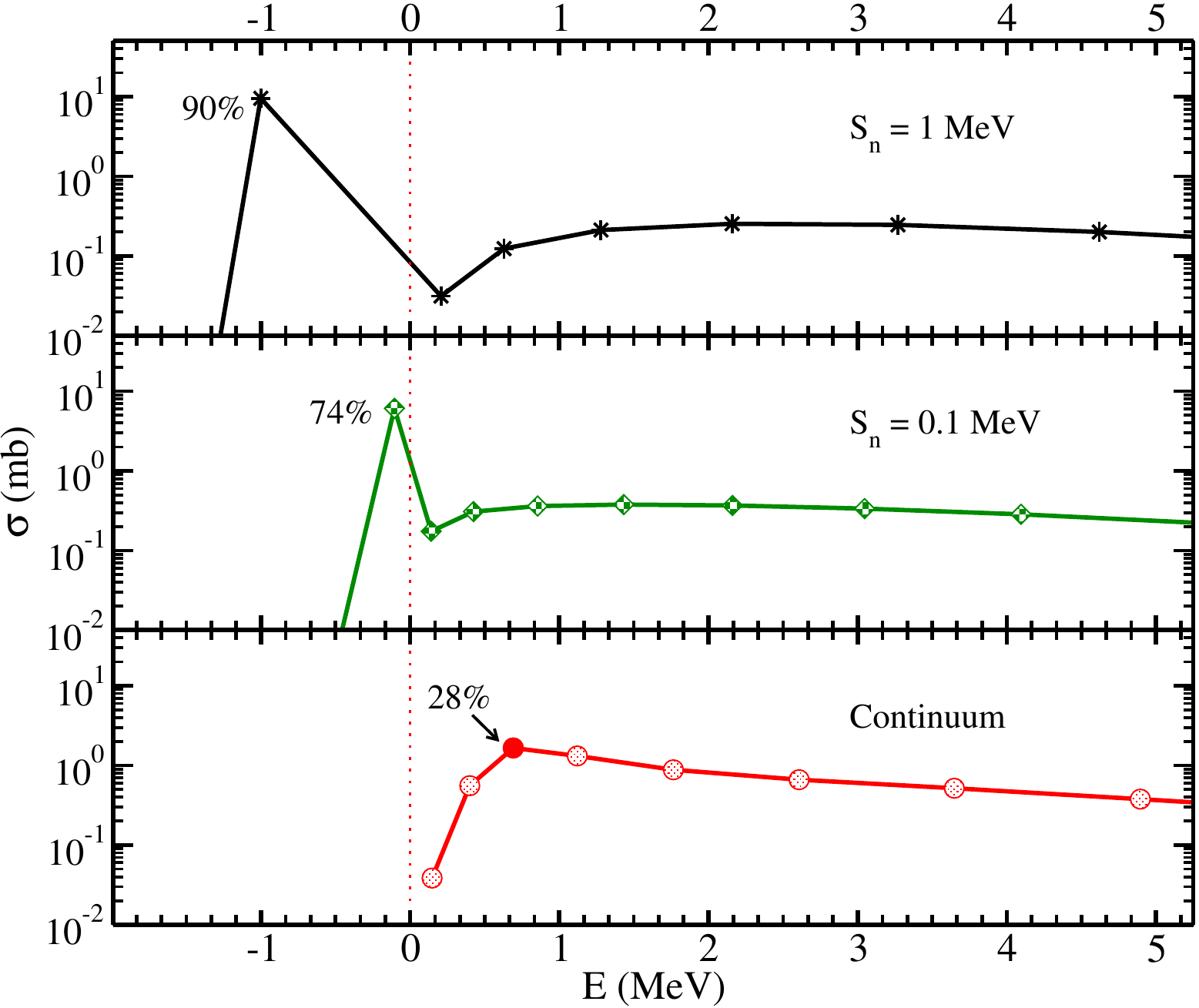}
	\caption{\label{fig: 1n_transfer} Total cross-section for one-neutron transfer during the $^{18}$O($^4$He,$^6$He)$^{16}$O reaction to form $^5$He in the three different scenarios. The threshold in $^5$He is marked at 0\,MeV with states above that representing the discretized continuum. The resonance state is displayed by the solid red circle for the continuum case. The percentage contributions of the highest contributing states are 90\%, 74\% and 28\%, respectively. For details, see text.}
\end{figure}

One-neutron transfer cross-sections for the system were computed thus, and are shown in Fig. \ref{fig: 1n_transfer} for the three scenarios. The results are displayed up to 5\,MeV energy. It is noticeable that the pseudo-states used to mimic the photography of $^5$He spectrum in each of the cases get separated more in energy as one goes higher in the spectrum.  The contributions of the bound states in the $S_n$ = 1\,MeV and $S_n$ = 0.1\,MeV cases are expectedly the largest. These states contributed about 90\% and 74\% in their respective system considerations, giving cross-section values of about 9.94\,mb and 6.12\,mb, respectively. The third state in the continuum case, being the resonant state (and shown with a solid red circle), was the most dominant state with a cross-section of 1.66\,mb, which was a contribution of about 28\% to the total one-neutron transfer cross-section.
With the spectra so generated and the one-neutron transfer showing the (expected) dominant state for each of the cases, we now proceed to have a look at the two-neutron transfer with an anticipation that these states should also dominate the two-neutron transfer canvas.





\subsection*{Two-neutron transfer}

An important aspect of such a two-particle transfer is the pairing interaction that causes the pairing between the particles being transferred. The inclusion of this interaction would involve adding it as a perturbation to an unperturbed Hamiltonian followed by diagonalization to obtain the required eigenvalues. For the perturbative part, one actually needs a residual interaction between the continuum states of the two particles in $^6$He \cite{FCS14PRC}. For the moment, a contact delta form, $\Delta = -g(\vec{r_1}-\vec{r_2})$ (with $g$ being the coupling strength in MeV-fm$^{-3}$) is used in describing the attractive pairing \cite{SFV22PLB}, although density dependent interactions might appear to be better suited \cite{BE91AP} as they can allow one to study BEC-BCS crossover effectively \cite{HSC07PRL}. The reason for using such a contact delta interaction for the pairing potential is the ease in numerical implementation (as this form has just one parameter that need be adjusted) \cite{SFV16EPJ,FCS14PRC} and sufficiency for the present purpose \cite{SFV22PLB}. This parameter has been adjusted throughout this work to modify the strength of the pairing interaction so as to reach different scenarios and consequently, different final states in the $^6$He final product.

With each subsequent transfer of a particle in a multi-particle transfer, the increase in the number of configurations that could potentially contribute towards the transfer process results in a combinatorics problem \cite{OV01RPP}. However, whatever be the total number of combinations\footnote{Depending upon the number of states involved, for $N$ states the number of combintaions would be $N(N+1)/2$. For example, here $N$ = 8 results in 36 different basis state configurations.}, for a two-particle transfer, all the combinations or `configurations' can be distinguished into three categories. For any given configuration $|p,q\rangle $ (with $p, q \in \{1, ... , N\}$), meaning that one particle is in state $p$ while the other is in state $q$, the system would be in the lowest energy when both the particles fill up the ground state (or the $|1,1\rangle $ state). Next in energy would be the class where one of the particles remains in the ground state, but the other goes in any of the excited states. This category would be represented by $|1,q\rangle $ (with $q \neq 1$). The third category is of all the configurations where both the particles are in any of the different excited states, or the $|p,q\rangle $ (with $p,q \neq 1$) configurations. Notice that all the $|p,p \rangle $ configurations form a pure configuration, where the particles pair in the same state with $|1,1\rangle $ being the lowest in energy (cf. Fig. 2 of Ref. \cite{SFV22PLB}). Then the probability of occupation of each of these configurations would be an ideal way to analyse the pairing effect in the two neutron transfer as the pure configurations should always have a higher occupation probability than the mixed configurations.


With this visualization, let us now see the behaviour of these basis state configurations for two-neutron transfer. One would expect the basis states with a substantial one-neutron transfer to contribute more to a subsequent neutron transfer as well.

\subsection{Varying Pairing potential}
	\label{sec:Delta}

I start by varying the pairing potential $\Delta$ and analysing the probabilitiy coefficients of the basis states for the properly bound case. The pairing potential has been varied from expected values like, 1\,MeV to a `seemingly' unphysical value of 17\,MeV to account for an extreme case at the other end of the spectrum.

\begin{figure}[htbp] 
	\centering
	\includegraphics[trim={0 0 0 0},clip,width=1.0\columnwidth]{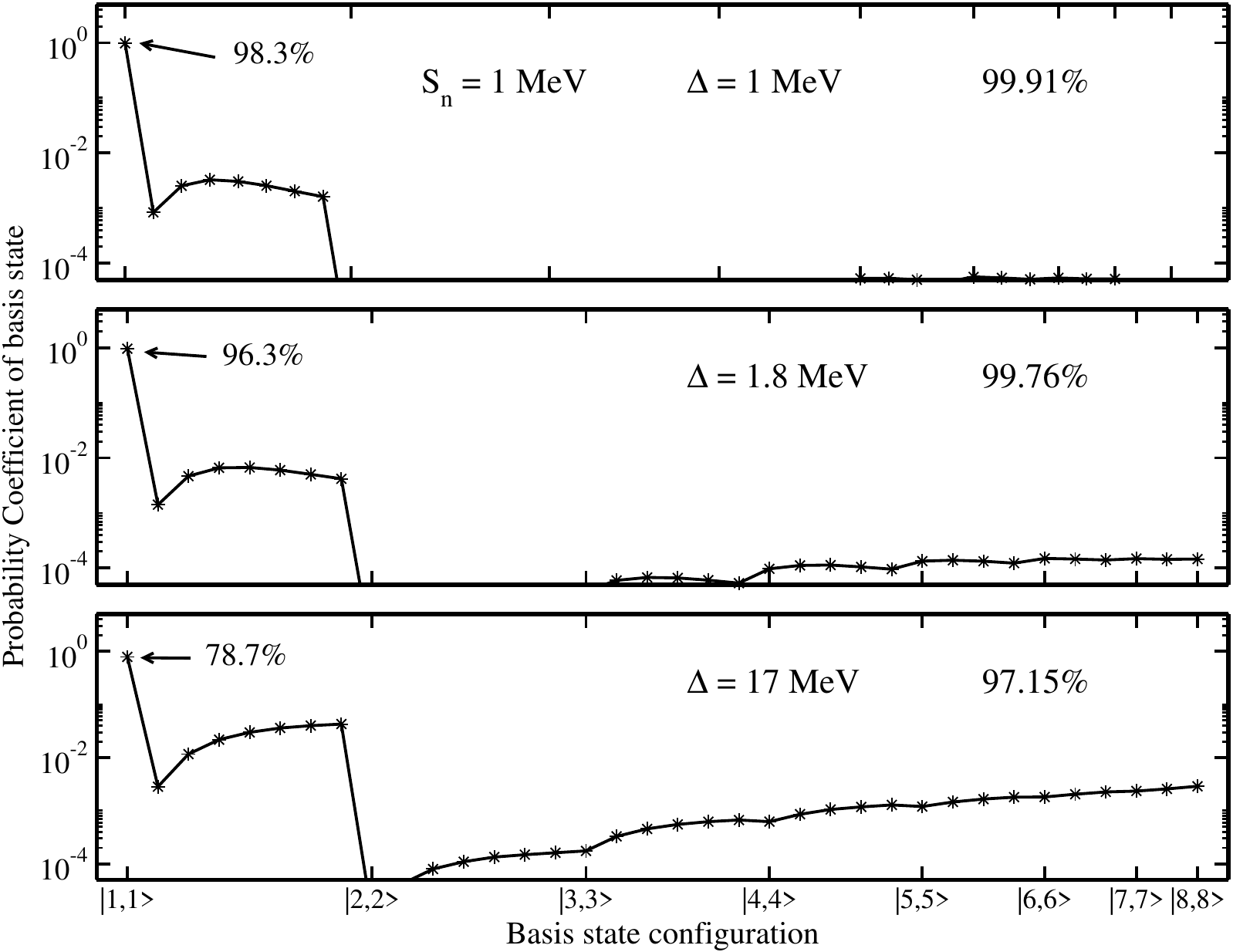}
	\caption{\label{fig: basis_contri_1MeV_DiffDelta} The probability contribution of each of the configurations through their respective bases for different strengths of the pairing interaction provided. The analysis is done when the intermediate $^5$He is considered bound by $S_n$ = 1\,MeV. As the pairing interaction increases, the contributions of the higher lying energy states increases, but ever so slightly.}
\end{figure}

{In Fig. \ref{fig: basis_contri_1MeV_DiffDelta}, one can see the probability coefficients corresponding to the contribution of each of the basis state configurations for varying pairing interactions. The specific values of the pairing interaction $\Delta$ chosen were obtained by fixing a value of the coupling constant $g$ that was used in Ref. \cite{SFV22PLB} for the $S_n$ = 0.1\,MeV as well as the continuum case (please see Table I in Ref. \cite{SFV22PLB}). This resulted , for the $S_n$ = 1\,MeV case, in $\Delta$ values of 1.8\,MeV and 17\,MeV, respectively. $\Delta$ = 1\,MeV was added for comparison. They will all lead to different final states in the $^6$He nucleus. The values of the coupling strength required to generate the desired pairing interaction and the corresponding final state in $^6$He are listed in Table \ref{T1}. The analysis in Fig. \ref{fig: basis_contri_1MeV_DiffDelta} is shown for the bound case of $^5$He with $S_n$ = 1\,MeV, where the properly bound nature of $^5$He demonstrates a weak dependence on probability coefficients responsible for pairing enhancement.}


\begin{table}[htbp]
		\caption{\label{T1} {Table showing the coupling constant strength $g$ (in MeV-fm$^{-3}$) required to produce the pairing interaction $\Delta$ (in MeV) mentioned for each case under study.}}
		\centering
		\vspace{0.50cm}
		\begin{tabular}{|*{5}{c|}}
			\hline\hline
			Case &$\Delta$ (MeV)  &  $-g$ (MeV-fm$^{-3}$) & $^6$He g.s. (MeV)  \\
			
			\hline
			 & 1& 576 & -3   \\
			$S_n$ = 1\,MeV & 1.8 & 992 \cite{SFV22PLB} & -3.8  \\
			 & 17 & 7827 \cite{SFV22PLB} & -15  \\
			\cline{1-4}
			Continuum & 1.8 & 6512 & -0.42  \\
			& 17 & 36798 & -15.62 \\
			
			\hline\hline
		\end{tabular}
\end{table}
	
One can clearly observe that with increasing the interaction, the pairing contribution for the $|1,1\rangle$ bound state configuration decreases. This is a result of the increased pairing contributions in the $|1,q\rangle$ ($q \neq$ 1) configurations meaning that the probability of particle interactions with at least one particle in the bound state increases due to the increase in the pairing potential. In other words, the particles gain energy to occupy higher energy levels. As can be seen from the lower most panel of the figure, it requires a very large amount of pairing interaction ($\Delta$ = 17\,MeV) to promote the particles to higher energy levels that are not directly generated via the bound state. Such a strong pairing potential is then sufficient to promote the particles to the energy levels of the continuum. Nevertheless, the large pairing energy required shows the reluctance of a closed quantum system to proceed for a reaction via its higher lying unbound energy states. Since 1\,MeV is not a very large separation energy for a nucleus, it just suggests that more tightly bound nuclei, when playing the role of an intermediate nucleus in a two-neutron transfer, should not contribute much to this pairing enhancement.

\begin{figure}[htbp] 
	\centering
	\includegraphics[trim={0 0 0 0},clip,width=1.0\columnwidth]{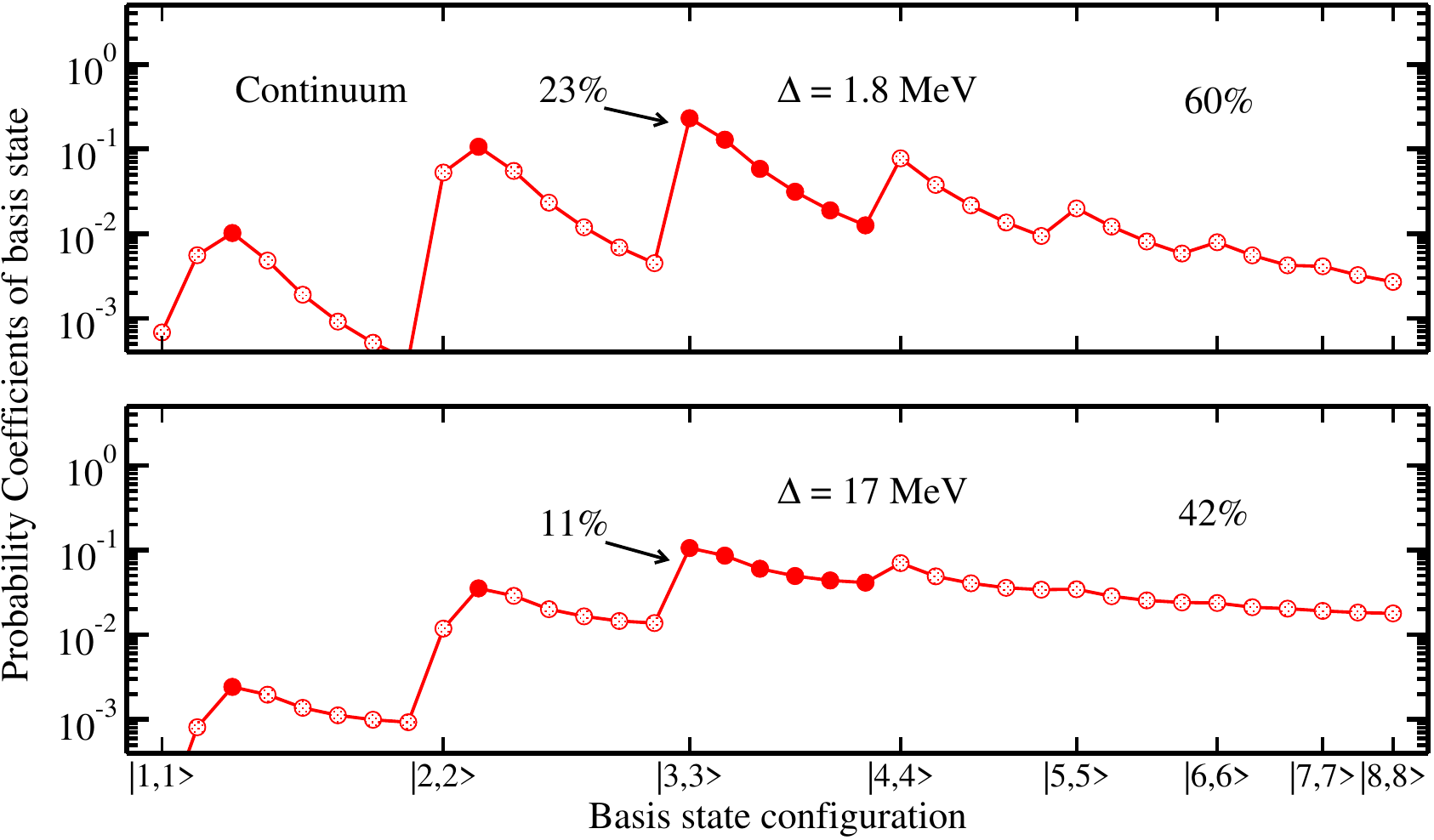}
	\caption{\label{fig: basis_contri_cont_DiffDelta} Same as Fig. \ref{fig: basis_contri_1MeV_DiffDelta}, but for the continuum case. The reduction in the percentage contribution of the resonance state (sum of solid circle percentages) due to increased pairing strength is clealy seen, going from 60\% to 42\%. It also throws light on the importance of the position of the resonance state in the spectrum. For details, please see text.}
\end{figure}

Fig. \ref{fig: basis_contri_cont_DiffDelta} then presents a similar analysis, but for the continuum case. However, since a pairing of 1\,MeV is insufficient to bring the resonance state of the generated continuum in $^5$He to form a bound state in $^6$He, it has not been included here and only two scenarios can be seen.

Much like the $S_n$ = 1\,MeV bound case, an increase in the pairing interaction leads to higher pairing in the higher lying states. As a matter of fact, what is noteworthy is that the pairing correlation has increased so much that the resonant state configuration $|3,3\rangle $, though still dominant, is forced to reduce its contribution - going from 23\% in the first case to 11\% in the lower panel - because the higher lying states become more and more prominent. Even the overall contribution of the resonant state (through all the $|3,q\rangle $ configurations shown by the solid red circles in the figure) decreases from 60\% to 42\%. This reduction is due to greater contributions of the higher lying states. The increased contributions can even be seen to jump by almost an order of magnitude in some cases (cf. the $|3,8\rangle $ configuration). However, the pure $|p,p\rangle $ configurations still dominate in comparison to the corresponding $|p,q\rangle $ configurations. One should then expect that with further increasing pairing interaction, the contributions of the pure and mixed configurations should tend towards becoming equal. Indeed, strong pairing correlations imply high excitation energy states in the intermediate nucleus \cite{OV01RPP}.

This further seems to suggest that the position of the resonance state also matters when dealing with a sequential two-neutron transfer via the continuum of the intermediate nucleus. If the resonance state lies too high in energy, the effect of the resonance may not be too profound to make a substantial difference as the lower lying states involved in the transfer might contribute more due to the reduced pairing strength. In other words, the pairing potential might not be strong enough to force the particles to reach the high lying resonance and thus, affect enhancement in the transfer. On the other hand, if the resonance state is close to the threshold, but the pairing potential is too high, even then it leads to a similar, fairly undesirable situation, the increased correlations increasing the contribution of higher lying states.\footnote{Since in the formalism the pairing potential is treated as a perturbation, having such a large perturbation in comparison to an unperturbed Hamiltonian might make things complicated for a realistic case. Apart from possibly leading to an involvement of the BEC-BCS crossover \cite{HSC07PRL,Matsuo06PRC}, it might promote simultaneous transfer terms to dominate rather than sequential transfers \cite{PIB13RPP}, but a sizeablle contribution from the latter would still be present and for the present academic study, does present interesting limiting sets of analysis.}

	
\subsection{Fixed Pairing potential}
\label{transfer}

Having studied the evolution of the pairing probabilities via the probability coefficients of the basis state configurations for different pairing potentials, it would now be intersting to fix the latter and see the effect on the probability coefficients as well as the two-neutron transfer cross-sections.

\begin{table}[ht]
		\caption{\label{T2} {Table showing the two-neutron transfer cross-sections (in $\mu$b) corresponding to the coupling constant strength $g$ (in MeV-fm$^{-3}$) for each case under study. The pairing interaction, $\Delta$ was kept constant at 1.4\,MeV to reproduce different hypothetical g.s. energies of $^6$He corresponding to each case. The number of basis states contributing, $N$, was fixed at 8. The unperturbed cross-sections $\sigma_{2n}^{(u)}$ ($\mu$b) are also indicated.}}
		\centering
		\vspace{0.50cm}
		\begin{tabular}{|*{11}{c|}}
			\hline\hline
			Case & $^6$He g.s. & $-g$ & $\sigma_{2n}^{(u)}$ & $\sigma_{2n}$  \\
			
			& (MeV) & (MeV-fm$^{-3}$) & ($\mu$b) &  ($\mu$b)  \\
			\hline
			$S_n$ = 1\,MeV & -3.4 & 784.5 &  445.42 & \color{red} 542.63 \\
			$S_n$ = 0.1\,MeV & -1.2 & 1228 & 115.67 & \color{red} 173.38  \\
			Continuum & -0.01  & 5472 & 4.63 & \color{red} 33.91  \\
			
			\hline\hline
		\end{tabular}
\end{table}

\begin{figure}[htbp]
		\centering
		\includegraphics[trim={0 0 0 0},clip,width=1.0 \columnwidth]{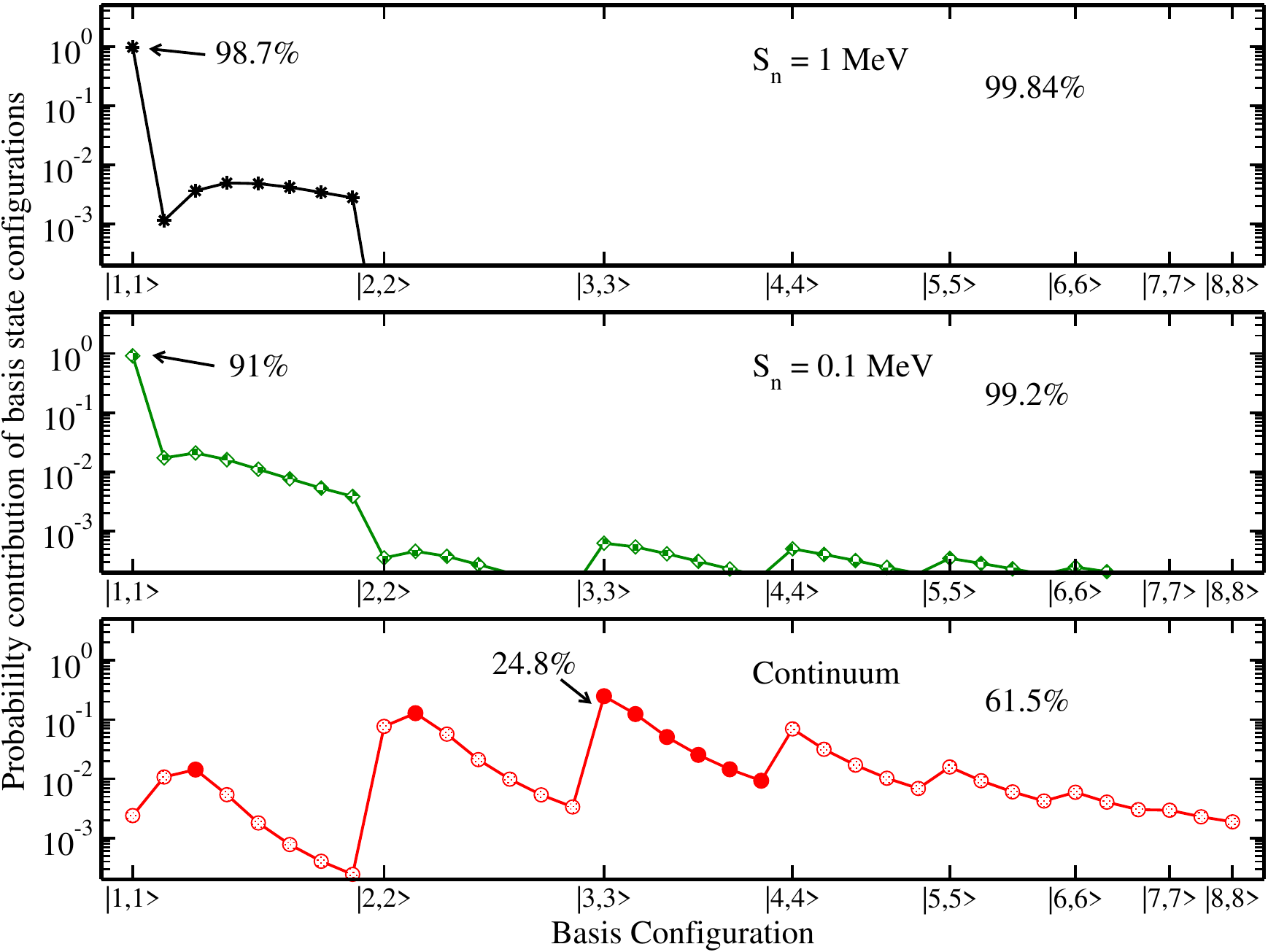}
		\caption{\label{fig: basis_fixedDelta} The probability coefficients for different configurations in the $^{18}$O($^4$He,$^6$He)$^{16}$O two-neutron transfer reaction through the respective bases for each of the three cases under study. All the $|p,q\rangle $ ($p,q \neq 1$) configurations in the bound state scenarios contribute negligibly while they make up almost all the contribution in the continuum case. The solid red circles serve as guide to the eye for configurations involving the resonance state. Further, as is evident the highest particle occupation probability is packed in the $|p,p\rangle $ configurations, except for the configurations in the continuum where one of the states is a resonance. For details, please see text.}
\end{figure}

Fixing the interaction potential $\Delta$ = 1.4\,MeV, Fig. \ref{fig: basis_fixedDelta} displays the probability coefficients for the occupation of levels configured by the various basis states under the three cases of study. $\Delta$ was fixed at 1.4\,MeV so as to enable the final state in $^6$He to be just bound at 0.01\,MeV for the continuum case. Then the $^6$He final states obtained for the two bound cases were at 1.2\,MeV and 3.4\,MeV corresponding to $S_n$ = 0.1\,MeV and $S_n$ = 1\,MeV, respectively (cf. Table \ref{T2}). 

It is seen in Fig. \ref{fig: basis_fixedDelta} that in the bound $^5$He cases, as was expected following the analysis of Figs. \ref{fig: 1n_transfer} and \ref{fig: basis_contri_1MeV_DiffDelta}, most of the contribution is due to the bound state, giving at least one of the particles more than 99\% chance of occupying this level. The individual contributions to the $|1,1\rangle $ bases are at 98.7\% and 91\% for $S_n$ = 1\,MeV and $S_n$ = 0.1\,MeV, respectively, decreasing with decreasing separation energy due to the increasing involvement of the continuum. In fact, the bound states are so dominant even in pairing that all the $|1,q\rangle $ configurations combine for occupation probabilities of 99.84\% and 99.2\% for the two cases, respectively, leaving negligible contribution of the continuum states to pair among themselves and contribute to the sequential transfer.

Yet, the picture is far more interesting when the intermediate $^5$He nucleus is unbound, as is the natural scene. Despite the presence of a resonance, all the continuum states tend to couple amongst themselves, resulting in a significant probability contribution for all the possible basis configurations. That the resonant state contributes the most is also expectedly observed, its pure contribution being about 24.8\%. Much like the bound state scenario, all the configurations with at least one resonant state in their makeup (solid red circles) contribute more than their respective pure $|p,p\rangle $ configuration, giving the total contribution of the resonant state to be about 61.5\%. Of course, apart from the configurations involving the resonance (like the $|2,3 \rangle $), all the pure pairing $|p,p\rangle $ configurations contribute more than their $|p,q\rangle $ configurations. However, the non-negligible contributions of all $|p,q\rangle $ configurations manifests that it is essential for a propoerly modelled continuum to be considered for such transfers where the particle pairings can move through different paths and interfere with each other according to the weights of these paths.

\subsection{The case of $^{22}$C}

{Let us now shift the analysis to a case where the system forms an even even Borromean halo, but with the presence of a virtual state in the intermediate system. Despite a paucity of such candidates, perhaps the best example is $^{22}$C forming due to the sequential addition of two-neutrons to $^{20}$C via $^{21}$C. $^{21}$C is unbound and supposedly has an $s$-wave virtual state above zero. The scattering length of the $s$-wave was obtained to be -2.77\,fm, in close proximity to the value of -2.8\,fm reported in Ref. \cite{Jagjit19FBS} and at 0.0288\,MeV, formed the first of the eight states in the discretized continuum.} 

There are several strategic advantages of choosing $^{22}$C. The g.s. energy of $^{22}$C is -0.140\,MeV \cite{GMO12PRL} and as such, it is very weakly bound, requiring only a pairing strength of 0.1688\,MeV to populate from the generated virtual state. For the $^{18}$O($^{20}$C,$^{22}$C)$^{16}$O reaction, 
two sets of calcualtions are done, both at 100\,MeV, not unlike the He case. Further, not only does choosing this reaction allow one to see the effect of pairing enhancement in the continuum with a virtual state (which is a significant difference than from a resonance state \cite{McVoy68NPA}), considering only the s-wave in the $^{21}$C continuum ensures that for an $^{18}$O projectile, we now have a case where the parity of the states in both the intermediate nuclei ($^{21}$C and $^{17}$O) remains the same. Morever, since the number of channels was drastically reduced due to the presence of $\ell$=0 angular momentum in considering the wave functions of both the intermediate nuclei, this allowed the inclusion of both the $d$- and the $s$-wave components in the g.s. wave function of $^{17}$O in the calculations.

The one-neutron transfer cross-section for this reaction was then found to be 33.46\,mb at 100\,MeV beam energy. 
A large part of these cross-sections were, of course, due to the s-wave transfer components. The s- to d- strengths were found to be orders of magnitude lower, and thus, negligible in comparison.


Moving to the two-neutron transfer part, it should be noted that since the energy gap required to pair the neutrons is very small going from the virtual state in the 21C continuum to the g.s. in $^{22}$C, which is at -0.140\,MeV, the coupling constant strength required to involke pairing was quite small. $g$ for this case came out to be just 0.03815\,MeV-fm$^{-3}$. Nevertheless, even this small pairing seems to play an important role when considering enhancement, as seen below.

\begin{figure}[htbp]
	\centering
	\includegraphics[trim={0 0 0 0},clip,width=1.0 \columnwidth]{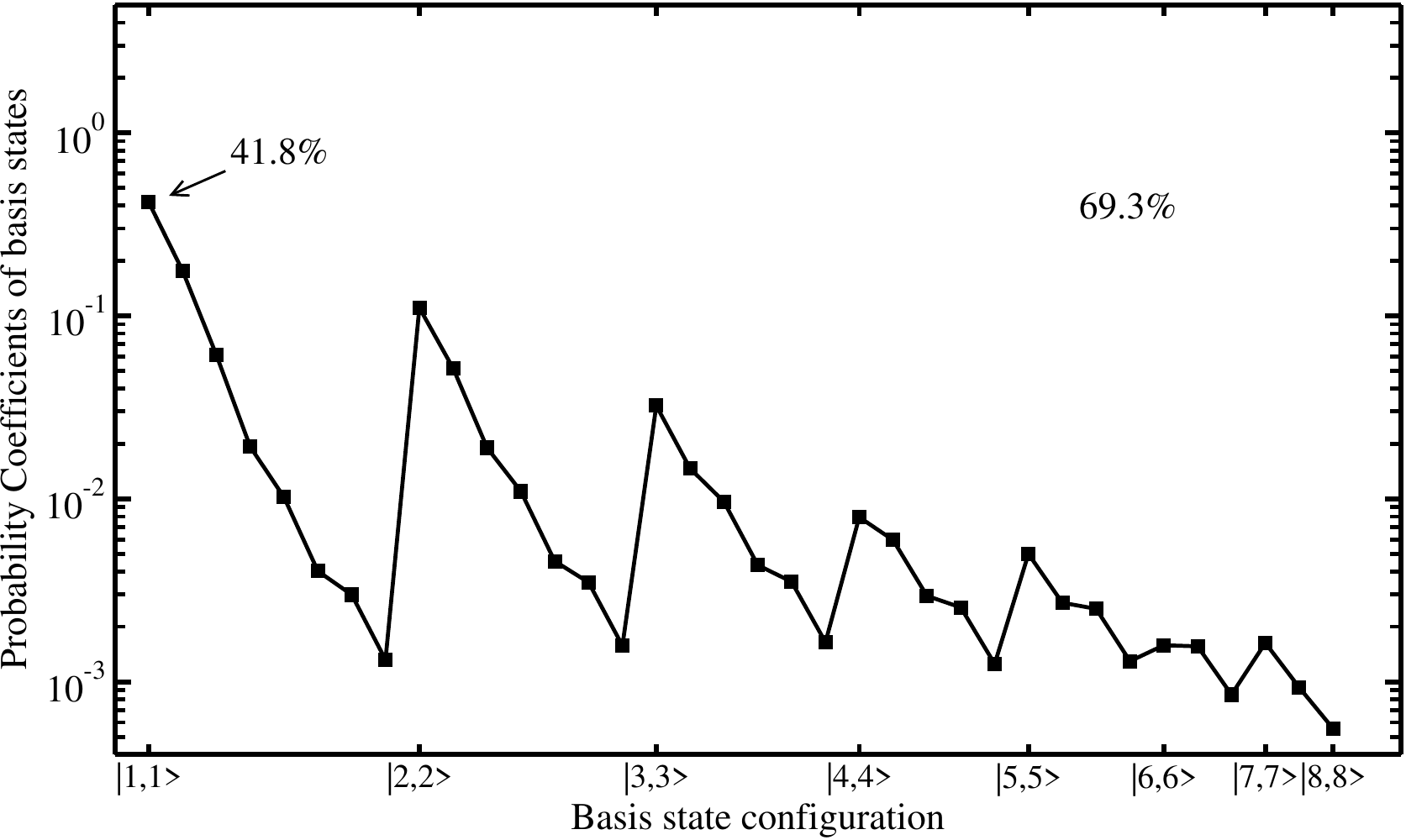}
	\caption{\label{fig: basis_21C} The probability coefficients for different basis state configurations in the $^{18}$O($^{20}$C,$^{22}$C)$^{16}$O two-neutron transfer reaction. Notice that the highest particle occupation probability is packed in the $|p,p\rangle $ configurations. For details, please see text.}
\end{figure}

In Fig. \ref{fig: basis_21C} we see the probability contribution of the basis states in the $^{18}$O($^{20}$C,$^{22}$C)$^{16}$O reaction forming $^{22}$C as the final product. It is evident that the pure $|p,p\rangle $ configurations are the ones that give most of the contribution towards neutron occupancy here, with $|1,1\rangle $ configuration (i.e., both particles in the virtual state), alone accounting for about 41.8\% of the total contribution. Overall, the $|1,p\rangle $ configurations have 69.3\% chance of being occupied. Even the contribution of configurations involving states lying closer to the threshold is higher than those involving higher lying states. For example, basis configuration $|1,4\rangle $ has a higher probaiilty coefficient than $|2,4\rangle $, which is higher than $|3,4\rangle $, that in turn dominates $|4,4\rangle $. Thus, contributions decrease with increasing energy in the continuum. 
The percentage decrease, of course, depends mainly upon the energy separation between the considered states.

\subsection{The cross-sections}

\begin{figure}[htbp]
	\centering
	\includegraphics[trim={0 0 0 0}, clip, width=1.0\columnwidth]{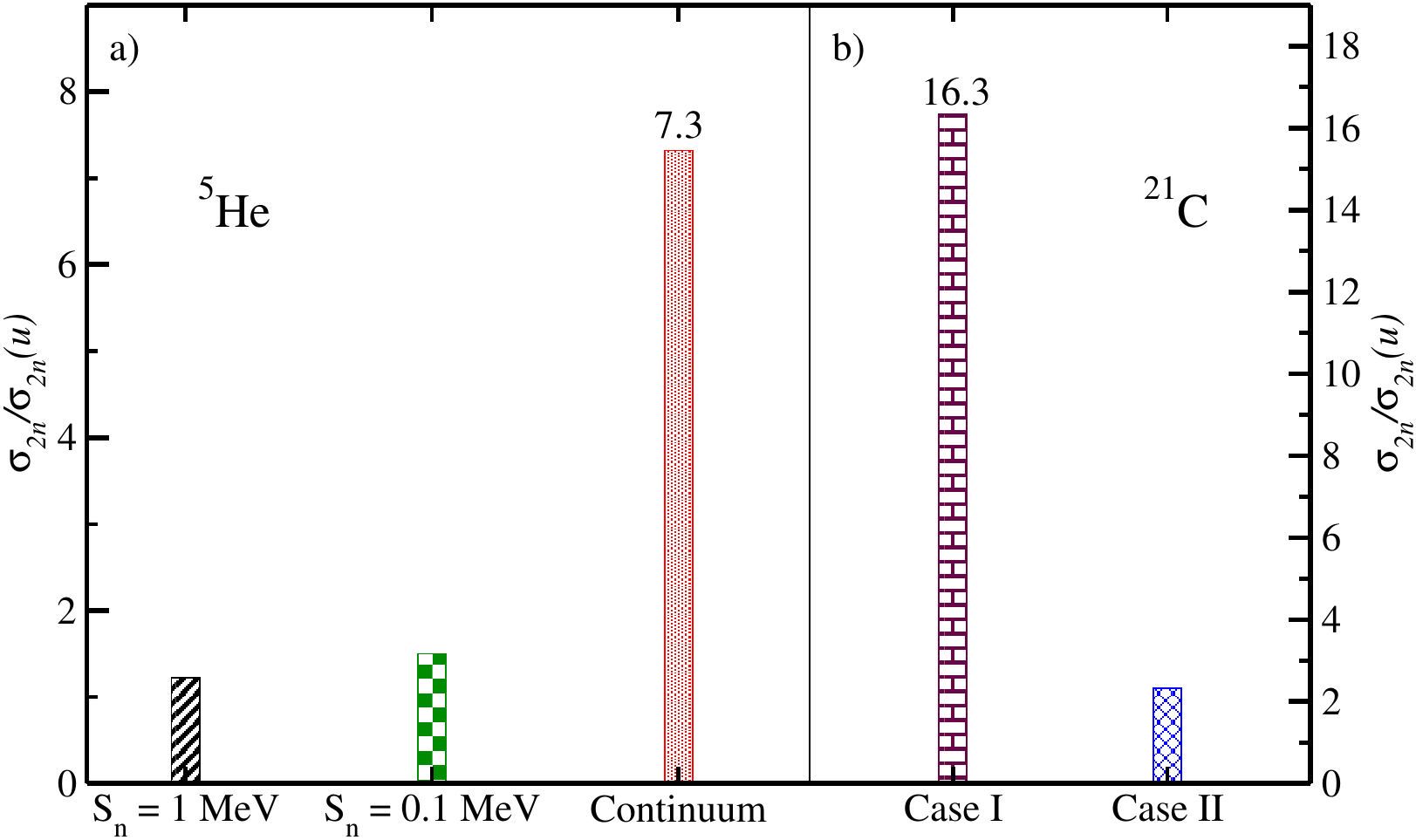}
	\caption{\label{fig: Sigma2nRatio} \(\left. a \right)\) The ratio of the two-neutron transfer cross-sections with pairing to the unperturbed two-neutron transfer cross-sections for each of the three cases of $^5$He involved in the present study. \(\left. b \right)\) The same ratio for the two cases in $^{21}$C. The large ratios in the cross-sections for the continuum case when $^5$He is naturally unbound and for Case I in $^{21}$C are clearly an evidence of the large pairing enhancement.}
\end{figure}

Finally, pairing enhancement is best revealed in the absolute cross-sections for two-neutron transfer. A comparison of the unperturbed cross-section $\sigma_{2n}^{(u)}$ ($\mu$b), with that of a cross-section obtained with the inclusion of pairing interaction $\sigma_{2n}$ ($\mu$b), would be ideal. Table \ref{T2} reveals the perturbed and unperturbed cross-sections computed for the different scenarios of $^5$He under study. The pairing interaction $\Delta$ was kept constant in this analysis and taken the same as in Sec. \ref{transfer}. Similarly, Table \ref{T3} presents these cross-sections for the $^{21}$C cases. One sees that the unperturbed cross-sections are always lower than those obtained with a pair field. However, what is interesting is the degree to which this increase is notable. To visualize this \textit{enhancement}, in Fig. \ref{fig: Sigma2nRatio}, I plot the ratios\footnote{An advantage of taking a ratio is also that any effects (or the lack thereof) due to the form of the pairing potential considered are washed off.} of the perturbed and unperturbed two-neutron transfer cross-sections displayed in Tables \ref{T2} and \ref{T3}, $i.e.$, $\sigma_{2n}/\sigma_{2n}^{(u)}$. 

	
It is evident from panel \(\left. a \right)\) that while the bound cases have an enhancement of 1.2 and 1.5 for $S_n$ = 1\,MeV and $S_n$ = 0.1\,MeV, respectively, it is the continuum case that catches the eye with a factor of 7.3. This essentially means that for weakly bound nuclei, the presence of a pairing field sigificantly increases the probability of a two-neutron transfer reaction, especially for an unbound intermediate ($A$+1) system. Indeed, enhancement factor (EF) depends on the interefernce of all the paths (or states) available for the final nucleus through the ($A$+1) nucleus \cite{Oert91PRC,OV01RPP}. A factor of 7.3 here suggests that the pairing field in this case induced seriously pronounced coherence. Although higher lying states contributed progressively lesser, nevertheless, they were equally crucial in ehancement observation and must not be discarded in any realistic calculation involving drip line nuclei. Their non-negligible contribution and the significant enhancement paves a way for bound ($A$+2) Borromean systems even when the ($A$+1) system is unbound.

Another aspect to note is that the pairing between bound and continuum states is not as strong as between two bound states or amongst continuum states \cite{Hama03PRC}. This is also very clearly seen considering that the bound $^5$He cases show very small enhancement due to pairing effects between the solitary bound states and the continuum states of their respective spectra. It points to the affinity of the system to prefer the bound states almost always as opposed to the continuum.


\begin{table}[ht]
	\caption{\label{T3} {Table showing the two-neutron transfer cross-sections (in mb) corresponding to the coupling constant strength $g$ of 0.03815 (in MeV-fm$^{-3}$) for the $^{18}$O($^{20}$C,$^{22}$C)$^{16}$O reaction. The pairing interaction, $\Delta$ was kept constant at 0.1688\,MeV to reproduce g.s. energy of $^{22}$C at -0.140\,MeV. The number of basis states contributing, $N$, was fixed at 8 and all of them contributed to the cross-sections involving pairing. The unperturbed cross-sections $\sigma_{2n}^{(u)}$ (mb) are indicated for the two cases when, I. The entire strength of the transfer goes through the virtual state (or the first state in the continuum) and II. All the states of the continuum have an equal probability of contribution.}}
	\centering
	\vspace{0.50cm}
	\begin{tabular}{|*{11}{c|}}
		\hline\hline
		Case & $\sigma_{2n}^{(u)}$ & $\sigma_{2n}$  \\
		 & (mb) &  (mb)  \\
		\hline
		
		I &  1.20 & \color{red} 19.61   \\
		II (equal contri. in unpert.) & 8.42 & \color{red} 19.61  \\
		
		\hline\hline
	\end{tabular}
\end{table}

{Let us now take a look at the enhancement (if any) that the Carbon study provides. In Table \ref{T3}, the cross-sections for the two-neutron transfer for the two cases are listed. Case I is the more expected scenario where the unperturbed sequential transfer propagates through the first discretised state in the continuum. The rest of the continuum does not contribute. Then, the pairing enhancement is seen to cause a very significant jump in the cross-section, raising it to 19.61\,mb.}

{However, since the virtual state is quite close to the particle emission threshold, and since many of the other generated states in the discretised continuum of $^{21}$C were quite close in the energy spectrum, I consider another extreme possibility: to consider all the states being occupied with equal probability. The contribution of a resonance free continuum in a two-neutron transfer is an open problem, therefore, as an approximation, let us assume that all the states contribute equally. Hence, Case II in Table \ref{T3} is for this diegesis. It is seen that the unperturbed cross-section then rises to 8.42\,mb.} 

Having the cross-sections brings us now to the Enhancement Factor. Refering to panel \(\left. b \right)\) of Fig. \ref{fig: Sigma2nRatio}, in Case I, a very significant enhancement is observed. This is for the simple reason that the pairing calculation involves all the possible configurations due to all the discretised states in the $^{21}$C continuum interacting via the pairing, but the unperturbed case has contribution only from one state which is the mimic of the virtual state. On the other hand, considering Case II, when all the pure states of the discretised continuum have an equal contribution for an unperturbed system, the EF drops to 2.33. What is noteworthy is that it still provides an enhancement. 

	
More importantly, the role of pairing is seen explicitly. The analyses from both the $^5$He and $^{21}$C studies show that pairing enhancement occurs no matter the contribution of the intermediate nucleus, although the degree of enhancement may vary. Laterally, this also points to the fact that the continuum contribution needs to be quantized.




\section{Summary and conclusions}
	\label{conclusion}
To summarize, I have tried to paint a simplistic pairing enhancement picture in a two-neutron transfer process by taking different photographs of the intermediate nucleus spectrum. 
The reactions used for the purpose were $^{18}$O($^4$He,$^6$He)$^{16}$O and $^{18}$O($^{20}$C,$^{22}$C)$^{16}$O, where the final ($A$+2) products of $^6$He and $^{22}$C were modeled as two neutrons in the orbitals of the intermediate $^5$He and $^{21}$C nuclei, respectively. The analysis was built upon three different hypotheses on the $^5$He system, two of which considered this ($A$+1) nucleus to be bound. These consisted of considering $^5$He to have hypothetical one-neutron separation energies at $S_n$ = 1\,MeV and $S_n$ = 0.1\,MeV and were compared with a more natural variation of an unbound $^5$He whose resonance state was generated at about 0.7\,MeV above the particle emission threshold. For the cases involving $^{21}$C as the intermediate nucleus, its continuum was discretized such that the virtual state was formed at 0.0288\,MeV above the threshold.

It was seen in detail that a movement across the particle threshold for a weakly bound nucleus renders the inclusion of a properly modeled continuum indispensable to grasp the dynamics of a two-neutron transfer. It is known that pairing enhancement originates from the interactive correlations among the particles or excitations which cause specific phase relations in two-step processes \cite{Oert91PRC,OV01RPP}, and the present work showed that such correlations are much more prevelant and crucial in the unbound intermediate nucleus. This lead to significant pairing enhancement of the overall two-neutron transfer cross-sections and in a way made these correlations somewhat responsible for the existence of ($A$+2) Borromean systems despite the ($A$+1) systems being unbound. The position of the resonance in such a continuum is also found to play a role. On the other hand, the affinity of the system to proceed for transfer via the bound states (if and when present) was also observed. However, it must be noted that the issue of the quantification of the contribution of the resonance free continuum in an unperturbed system remains an open problem and needs to be tackled to establish with any certainty the effect the pairing enhancement can really have in a two-neutron transfer. Further, it will certainly be interesting to see the effects of this pairing enhancement in the neutron rich medium mass region with the inclusion of configuration mixing. Two-neutron transfers are a popular method to produce medium and heavier exotic nuclei and could lead to the discovery of heavier Borrowmean systems. The strength of the pairing leads to significant phenomenon like the BEC-BCS crossover \cite{PBV10PRL}, or influence the dynamic fission barriers in super heavy nuclei \cite{SDN14PRC}, and the analysis of the same from a reaction perspective under the present approach is something to look forward to for the future.\\

\bigskip
\section*{Acknowledgements}
This work is dedicated to Andrea Vitturi and his memory. I thank him for introducing me to the world of two-neutron transfers over a cup of cappuccino on a bright Wednesday morning in September 2021 outside the Dipartimento di Fisica e Astronomia 'G. Galilei', Padova. He was one of the finest humans I ever met, extremely kind and empathetic, but with very peircing eyes. He was and is an inspiration. I deeply thank Lorenzo Fortunato and Enrico Vigezzi for the insightful and critical discussions about the work and the manuscript. I am grateful to Antonio M. Moro for the patient and detailed clarifications regarding the THO basis.
I would also like to acknowledge Pierre Capel and PRISMA+ (Precision Physics, Fundamental Interactions and Structure of Matter) Cluster of Excellence, Johannes Gutenberg University Mainz for providing the computational facilities.


	
	
\bibliography{gaganbiblio}
	
\end{document}